\documentclass[runningheads]{llncs}

\usepackage[utf8]{inputenc}
\usepackage[T1]{fontenc}
\usepackage{amsmath,amssymb}
\usepackage{graphicx}
\usepackage{booktabs}
\usepackage{multirow}
\usepackage{array}
\usepackage{tabularx}
\usepackage{url}
\usepackage{xcolor}
\usepackage{hyperref}
\usepackage{cite}
\usepackage{balance}
\usepackage{tikz}
\usetikzlibrary{shapes.geometric,arrows.meta,positioning,backgrounds,fit}
\usepackage{pgfplots}
\pgfplotsset{compat=1.18}
\usepackage{enumitem}
\setlist{itemsep=1pt,topsep=3pt,parsep=0pt,partopsep=0pt}

\definecolor{colLSG}{RGB}{0,45,98}
\definecolor{colGEDES}{RGB}{0,100,55}
\definecolor{colACC}{RGB}{184,49,47}
\definecolor{colGRIS}{RGB}{100,100,100}
\definecolor{colBG}{RGB}{240,246,255}
\definecolor{colBGg}{RGB}{240,248,240}

\hypersetup{
  colorlinks=true,
  linkcolor=colLSG,
  citecolor=colLSG,
  urlcolor=colGEDES,
  bookmarks=true
}

\title{Signal-Driven Pervasive Game Design: The LifeSync-Games Framework as a Player Experience Integration Layer}

\author{Joaquín Macías-Cáceres\inst{1} \and
Francisco Gutiérrez-Vela\inst{2} \and
Patricia Rodríguez\inst{2} \and
Roberto González-Ibáñez\inst{1}}

\authorrunning{J. Macías-Cáceres et al.}

\institute{Universidad de Santiago de Chile (USACH), Santiago, Chile\\
\email{joaquin.macias@usach.cl}, \email{roberto.gonzalez.i@usach.cl}\\ \and
Universidad de Granada (UGR), Granada, España\\
\email{fgutierr@ugr.es}, \email{patricia@ugr.es}}

\titlerunning{Signal-Driven Pervasive Game Design: The LSG Framework}

\begin{document}
\maketitle
\begin{abstract}
Pervasive games extend the magic circle across spatial, temporal, and social dimensions, yet treat the player's physiological and cognitive state as a passive receptor rather than an active signal. This paper presents LifeSync-Games (LSG), a framework that (unlike proposals treating player signals as an additional dimension), operationalizes them as a Player Experience Integration Layer (PEIL) acting transversally across the three existing pervasive dimensions through verified real-world signals: physical activity, sleep quality, memory, and decision speed. The framework introduces a gamified integration artifact (the LSG portal) that mediates the real $\leftrightarrow$ virtual exchange through redeemable points, real-world missions, and structural gamification. Five HCI design principles grounded in Self-Determination Theory and Flow Theory are proposed, instantiated across six commercial video games, together with a study protocol ($n \approx 70$-80 participants, quasi-experimental design). \textbf{This paper reports the design stage of LSG}: rule thresholds and portal parameters are design decisions pending empirical calibration; no data collection has yet been conducted. The main contribution is a theoretically grounded framework and validation protocol positioning player-sensitive integration as the mechanism enabling pervasive games to respond to the player's actual biological and cognitive state.
\end{abstract}


\section{Introduction}
\label{sec:intro}
Pervasive games \cite{montola2005} extend the magic circle across spatial, temporal, and social dimensions, yet treat the player's physiological and cognitive state as a passive receptor rather than an active signal capable of reconfiguring the game itself. This gap is not merely conceptual: existing pervasive game frameworks \cite{medina2021,arango2021geopgd} adapt to \textit{where}, \textit{when}, and \textit{with whom} the player acts, but not to the player's actual biological and cognitive condition. Health-oriented gamification systems attempt to close this gap, but do so unidirectionally and symbolically -- real-world behaviour is logged and rewarded with decorative badges rather than reconfiguring the game itself, and adherence collapses once novelty fades \cite{hamari2014,johnson2016,zhao2024}. This leaves an open problem: how can verified physiological and cognitive signals be integrated \textit{into} game mechanics, across established pervasive dimensions, without coercion or redefining the pervasive ontology? This paper presents \textbf{LifeSync-Games} (LSG), a framework that answers this problem by operationalizing the player's real-world state not as an additional dimension to Montola's model, but as a \textbf{transversal layer} that instruments the three existing dimensions through verified physiological and cognitive data. The main contributions are:

\begin{enumerate}
    \item The conceptualization of LSG as a \textbf{Player Experience Integration Layer} (\textbf{PEIL}) that transversally instruments Montola's three dimensions \cite{montola2005} through verified signals, as opposed to treating them as an independent parallel dimension.
    \item A \textbf{gamified integration artifact} (the LSG portal) that mediates the real $\leftrightarrow$ virtual exchange through redeemable points, real-world missions, and structural gamification.
    \item Five \textbf{HCI design principles} for signal-driven pervasive games, grounded in SDT \cite{deci1985,ryan2000} and Flow Theory \cite{csikszentmihalyi1990}.
    \item A \textbf{game mechanics mapping catalogue} instantiated across six commercial video games.
    \item A \textbf{study protocol} with ethical approval.
\end{enumerate}

This work extends the evaluation framework of Medina-Medina et al. \cite{medina2021} and the GeoPGD methodology \cite{arango2021geopgd}. The paper is organized as follows: Section 2 reviews related work; Section 3 presents the LSG framework (PEIL Sect. 3.1, pervasive dimensions Sect. 3.2, HCI principles Sect. 3.3, LSG portal Sect. 3.4, pipeline Sect. 3.5); Section 4 details the mechanics mapping catalogue; Section 5 describes the study protocol; Sections 6 and 7 present the discussion and conclusions.

\section{Related Work}
\label{sec:related}

This section establishes the theoretical and empirical foundations of LSG: Montola's pervasive dimensions as the conceptual basis (Sect. 2.1), the foundations of structural gamification and its limitations in health settings (Sect. 2.2), and the motivational and contextual adaptation frameworks underpinning the design principles (Sect. 2.3).

\subsection{Pervasive Games and the Pervasive Dimensions}

Montola's foundational definition \cite{montola2005} extends Huizinga's classical notion of the \textit{magic circle}, the conceptual and social boundary that separates play from ordinary life, by arguing that pervasive games deliberately blur this boundary along three axes: \textbf{spatial} (no fixed location), \textbf{temporal} (no fixed sessions), and \textbf{social} (outsiders can influence the game). Unlike traditional games, where crossing the circle's edge ends the game, pervasive games are designed so the edge itself becomes permeable, a framework operationalized in heuristic evaluation \cite{medina2021}, design methodologies \cite{arango2021geopgd}, and applications in health and education \cite{arango2017review}. Medina-Medina et al. \cite{medina2021} hint at a potential additional axis (dynamic pervasivity), without formalizing it as a first-order dimension. LSG builds on this intuition by proposing a transversal mechanism that makes the three existing dimensions responsive to the player's verified real state. LSG's positioning also differs from two established families of player-experience integration. Dynamic Difficulty Adjustment (DDA) systems \cite{mortazavi2024} sense in-session performance or affect and adjust challenge in real time, but the loop closes within a single session; affective and biofeedback-driven games \cite{bontchev2016} extend this to physiological signals, yet remain single-game, single-session systems without a cross-game mediating artifact. LSG differs in three respects: it operates on verified, non-self-reported signals accumulated \textit{between} sessions rather than in-session telemetry; it mediates the exchange through an external artifact (the LSG portal) decoupled from any single title; and it instruments Montola's three pervasive dimensions transversally, rather than optimizing a single session's challenge-skill balance.

\subsection{Gamification: Foundations in Non-Game Systems and Health Applications}

Gamification, understood as the application of game design elements in non-game contexts \cite{deterding2011}, improves motivation when its elements are integrated into the system's usage flow \cite{hamari2014}; Seaborn and Fels \cite{seaborn2015} distinguish superficial gamification (achievements unrelated to use) from structural gamification (achievements anchored in authentic behaviours), and Nacke and Deterding \cite{nacke2017} argue that the most effective systems form a micro-world with transparent rules and user autonomy. In health settings, however, these conditions are systematically unmet: the relationship between real-world behaviour and game elements is unidirectional and symbolic \cite{hamari2014,johnson2016,zhao2024}, and the motivational mechanism collapses once novelty fades \cite{johnson2016}. LSG addresses this gap through (a) signal $\leftrightarrow$ mechanics integration, where the player's state directly modifies game parameters; and (b) structural gamification in the LSG portal, whose achievements are anchored to verified signals, meeting the conditions set by \cite{nacke2017,seaborn2015}.

\subsection{Self-Determination Theory, Flow, and Contextual Adaptation}

Self-Determination Theory (SDT) \cite{deci1985,ryan2000} posits that intrinsic motivation requires satisfying three psychological needs, \textbf{autonomy, competence, and relatedness}, and predicts that calibrating challenge to the player's actual state yields more sustained motivation than treating difficulty as independent from the real world \cite{ryan2006}. Flow Theory \cite{csikszentmihalyi1990} complements SDT: optimal experience emerges when challenge and skill are in balance. LSG extends this logic to video games by treating their parameters as continuous functions of the player's real physiological-cognitive profile.

\section{LifeSync-Games: The Player Experience Integration Layer}
\label{sec:framework}

This section presents the LSG framework following a progressive logic: PEIL positioning and pervasive instrumentation (Sects. 3.1-3.2), HCI design principles (Sect. 3.3), the LSG portal with its architecture and metrics (Sect. 3.4), and the real\,$\leftrightarrow$\,virtual synchronization pipeline (Sect. 3.5). Reading Sects. 3.1-3.3 establishes the conceptual foundation required to interpret the design decisions in Sects. 3.4 and 3.5.

\subsection{Conceptual Framework: From Dimension to Transversal Layer}

Montola's three dimensions \cite{montola2005,medina2021} break specific boundaries of the magic circle: spatial pervasivity eliminates fixed locations, temporal pervasivity eliminates fixed sessions, and social pervasivity eliminates the predefined player group. LSG does not propose a fourth boundary; instead, it introduces a transversal mechanism that makes the three existing dimensions responsive to the player's verified real state: physiological and cognitive signals do not create a new type of pervasivity, but rather instrument the three already existing ones (Table \ref{tab:dimensions}). The distinction is one of nature, not quantity: LSG endows those dimensions with a \textit{verified sensory substrate}, such that a player who walks simultaneously enriches spatial, temporal, and social pervasivity. Concretely, raw signals are normalized and combined into two per-dimension composite indices, the physical composite index ($\mathrm{IC}_{phys} \in [0,1]$) and the mental composite index ($\mathrm{IC}_{ment} \in [0,1]$), whose geometric mean produces the single LSG composite index ($\mathrm{IC_{LSG}} \in [0,1]$) that drives the rule catalogue (Sect.~\ref{sec:catalogue}) and the LSG portal (Sect.~3.4). To illustrate how this layer operates, the following section details the instrumentation mechanisms before describing the artifact that materializes them.

\begin{figure}[ht]
  \centering
  \includegraphics[width=0.9\columnwidth]{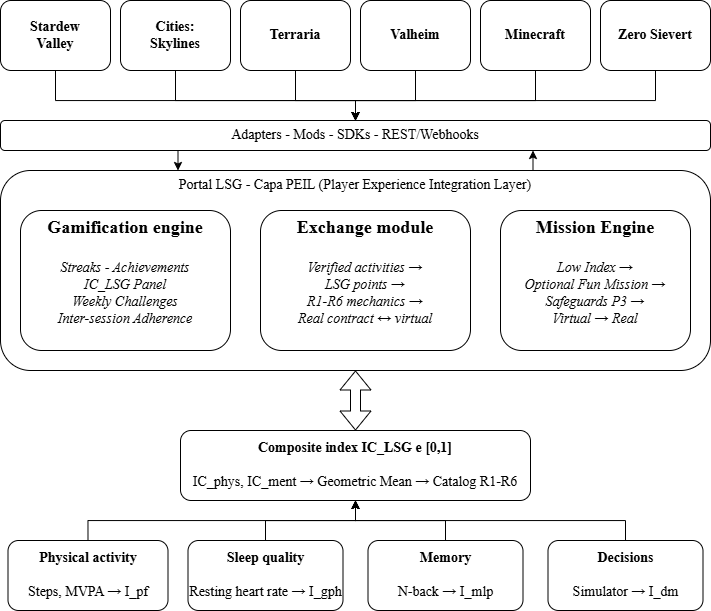}
    \caption{Conceptual framework of LifeSync-Games (LSG) as a Player Experience Integration Layer (PEIL). Real-world signals (bottom layer) are normalized into the composite index $\mathrm{IC_{LSG}}$ (aggregation layer), which feeds the LSG Portal and its three functional modules, Gamification Engine, Exchange Module, and Mission Engine (portal layer, detailed in Sect.~3.4), bidirectionally connected to the six video games (top layer) via adapters. Upward arrows indicate the real $\rightarrow$ virtual flow (verified signals driving game mechanics); downward arrows indicate the virtual $\rightarrow$ real flow (game events feeding back into the player profile).}
  \label{fig:lsg-marco-conceptual}
\end{figure}

\begin{table}[ht]
\caption{LSG as a transversal layer over Montola's three dimensions~\cite{montola2005}. The bottom row describes the integration mechanism, not an additional dimension. MVPA: moderate-to-vigorous physical activity.}
\label{tab:dimensions}
\centering
\footnotesize
\renewcommand{\arraystretch}{1.1}
\begin{tabular}{p{2.0cm}p{2.2cm}p{3.6cm}p{2.8cm}}
\toprule
\textbf{Dimension} & \textbf{Boundary} & \textbf{LSG Signal (rule)} & \textbf{Direction}\\
\midrule
Spatial & Fixed location & Physical activity\newline (steps, MVPA) $\rightarrow$\newline
  events linked to\newline real-world displacement (R1) & Real $\rightarrow$ virtual\\
Temporal & Fixed sessions & Sleep quality and\newline recovery $\rightarrow$\newline
  accumulated inter-session\newline modifiers (R1, R2) & Real $\rightarrow$ virtual\\
Social & Fixed groups & Group $\mathrm{IC_{LSG}}$ $\rightarrow$\newline
  cooperative and collective\newline narrative events (R6) & Real $\rightarrow$ virtual\\
\midrule
\textbf{PEIL Layer\newline (LSG)} & \textbf{Experiential\newline mediation} & \textbf{$\mathrm{IC_{LSG}}$ $\rightarrow$\newline
  point exchange +\newline real-world missions\newline
  via LSG portal (R1--R6)} & \textbf{Bidirectional\newline (Real $\leftrightarrow$ virtual)}\\
\bottomrule
\end{tabular}
\end{table}

\subsection{LSG and the Pervasive Dimensions: Concrete Instrumentation}

Signal-driven pervasivity shares properties with the three established dimensions (it operates continuously like the temporal dimension, and requires real-world interaction like the spatial one), but introduces a unique characteristic: it is the \textit{only} dimension that requires verified, quantified, and non-self-reported data from the player's body and mind. This turns the ethical imperatives of transparency, data minimization, and non-coercive safeguards into first-order design requirements. Table~\ref{tab:dimensions} summarizes the instrumentation mechanism, signal, and activated rule(s) per dimension: physical activity ($IC_{phys}$, R1) instruments spatial pervasivity, sleep and recovery ($I_{gph}$, R1--R2) instrument temporal pervasivity, and group $IC_{LSG}$ (R6, e.g., in Minecraft) instruments social pervasivity.

The design of these mechanisms is not arbitrary: it responds to five HCI design principles governing both the LSG portal and the game adapters, defined below and explicitly referenced in the architecture (Sect. 3.4.1) and the mechanics catalogue (Sect. 4).

\subsection{HCI Design Principles}
\label{sec:principles}

Grounded in SDT~\cite{deci1985,ryan2000} and Flow Theory \cite{csikszentmihalyi1990}, we propose five principles that extend the pervasive game design vocabulary \cite{medina2021,arango2021geopgd} to encompass physiological and cognitive signal integration; they apply to both the game and the LSG portal.

\begin{enumerate}
    \item \textbf{P1 - Transparent Mapping (autonomy):} Every rule mapping a signal to a mechanic must be visible and comprehensible to the player \cite{deci1985}, preserving agency; LSG's control panel displays index values, active rules, and modifier expiration times.
    \item \textbf{P2 - Calibrated Moderation (competence):} Signal-derived modifiers must be incremental, temporary, and bounded: LSG caps them at $\pm$15\% of base parameters for 24-96 h, operationalizing the SDT competence need.
    \item \textbf{P3 - Safeguard Design (non-coercive health signalling):} When signals indicate a health concern, the portal (not the game) offers optional missions framed as ludic challenges (R4-R5), preserving the fiction contract.
    \item \textbf{P4 - Bidirectional Narrative Integration (immersion):} Virtual $\leftrightarrow$ real synchronization must be integrated into the game's narrative \cite{arango2017review}, e.g., activity-induced pauses framed as \textit{seasonal farm events} in Stardew Valley.
    \item \textbf{P5 - Differential Pervasivity (player-controlled scope):} Players choose which signal dimensions participate in synchronization, addressing heterogeneity in privacy, health status, and hardware access.
\end{enumerate}

With the principles established, the following section describes the artifact that operationalizes them: the LSG portal, its architecture, and evaluation metrics.

\subsection{The LSG Portal: Gamified Integration Artifact}

Its function is threefold \cite{hamari2014,nacke2017,zhao2024}: (a) \textbf{Intrinsic gamification}, maintaining inter-session adherence through streaks, achievements, an $\mathrm{IC_{LSG}}$ panel, and weekly challenges; (b) \textbf{Real $\rightarrow$ virtual exchange}, converting verified activities into redeemable points exchangeable for game mechanics, with full traceability; and (c) \textbf{Real-world missions}: when $\mathrm{IC_{LSG}}$ falls below critical thresholds, the portal generates an optional mission (e.g., \textit{walk 20 min.\ $\rightarrow$ unlock a seasonal event in Stardew Valley}), preserving the fiction contract (P3). These three modules are not independent, parallel consumers of $\mathrm{IC_{LSG}}$: they interact through the shared Ledger (L3, Sect.~3.4.1). The Gamification Engine takes rolling admissibility and streak history as input and outputs redeemable points credited to the Ledger plus UI state (panel, streaks, challenges); it never interacts with the games directly. The Mission Engine takes threshold breaches on $\mathrm{IC}_{phys}$/$\mathrm{IC}_{ment}$ (R4, R5) as input and outputs an optional portal mission; on completion, it credits points to the Ledger. The Exchange Module takes the Ledger's point balance and a verified activity record as input, and outputs an R1-R6 mechanic unlock dispatched to L5, with a full traceability entry written back to the Ledger -- the only module with a direct connection to the six video games.

\subsubsection{LSG Portal Architecture}

The LSG architecture is structured into five layers L1-L5 (Fig.~\ref{fig:lsg-arquitectura}):

\begin{figure}[ht]
  \centering
  \includegraphics[width=0.9\columnwidth]{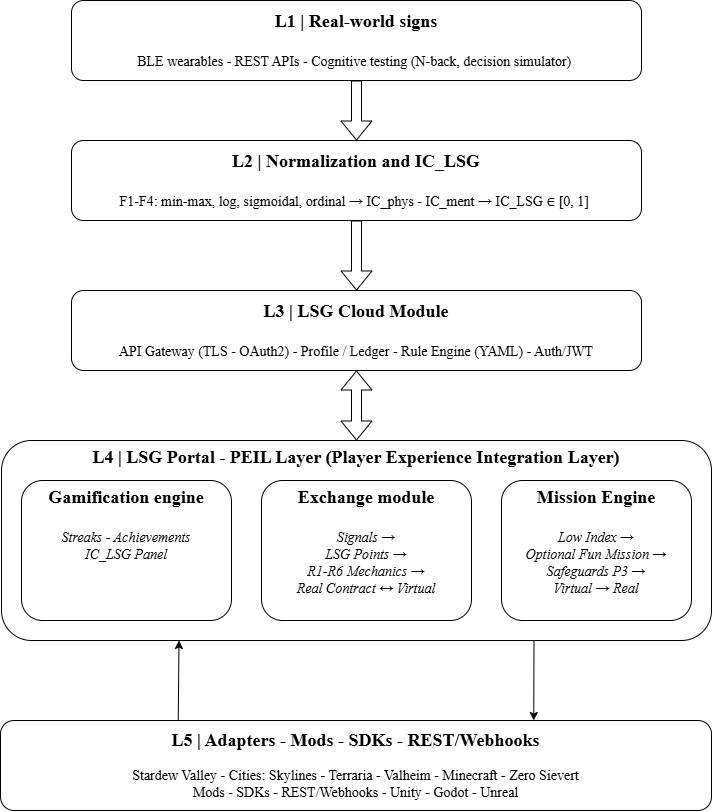}
  \caption{LSG portal architecture as the PEIL layer. Five layers (L1-L5) from real-world signals (L1, bottom) to adapters and video games (L5, top). L4 modules materialize the three functional flows of the portal: gamification, real $\leftrightarrow$ virtual exchange, and real-world missions.}
  \label{fig:lsg-arquitectura}
\end{figure}

\begin{enumerate}
    \item \textbf{L1. Real-world signals:} BLE sensors, REST APIs, and digital cognitive tests provide raw data; the layer verifies admissibility ($\geq\!70\%$ of daily valid samples) and applies conservative imputation.
    \item \textbf{L2. Normalization and $\mathrm{IC_{LSG}}$:} Strategies F1--F4 (min-max, log, sigmoidal, ordinal) normalize signals to $[0,1]$: F2 for the right-skewed $I_{pf}$; F4 for the Likert-based PSQI composite $I_{gph}$; F3 for the ceiling/floor-bounded $I_{mlp}$; and F1 for the already-scaled $I_{dm}$. The geometric mean of $I_{pf}$/$I_{gph}$ produces $\mathrm{IC}_{phys}$, of $I_{mlp}$/$I_{dm}$ produces $\mathrm{IC}_{mental}$, and of both produces $\mathrm{IC_{LSG}} \in [0,1]$.
    \item \textbf{L3. LSG cloud module:} API Gateway (TLS, OAuth2/JWT, p95 $\leq\!250$ ms), Profile/Ledger, Rule Engine (R1-R6 in declarative YAML), and Auth/JWT with per-sensor/per-game scopes. Pseudonymization (LSG-PXXX) separates identity from the analytical profile.
    \item \textbf{L4. LSG Portal -- PEIL Layer:} the Gamification Engine, Exchange Module, and Mission Engine described above (Sect.~3.4), cooperating through the Ledger.
    \item \textbf{L5. Adapters and video games:} SDKs and mods consume the cloud API via REST/Webhook; a local cache tolerates latency and game events generate callbacks (virtual $\rightarrow$ real), closing the loop.
\end{enumerate}

\subsubsection{LSG Portal Evaluation Metrics}

LSG portal effectiveness is measured across three dimensions: gamification, user interface, and commitment compliance (Table \ref{tab:metricas_portal}).

\begin{table}[h]
\caption{LSG portal measurement plan. Three dimensions: gamification (\textit{G}), interface (\textit{I}), and commitment compliance (\textit{C}). T0: baseline; T1-T4: intervention; T5: post-intervention. LSG-CV: LSG-linked condition (\textit{Con Vinculación}), i.e., LSG-integrated gameplay; LSG-SV: standard condition (\textit{Sin Vinculación}), i.e., gameplay without signal integration.}
\label{tab:metricas_portal}
\footnotesize
\begin{tabular}{llllll}
\hline
\textbf{Code} & \textbf{Dim.} & \textbf{Instrument} &
\textbf{Time-point} & \textbf{Threshold}\\
\hline
\multicolumn{6}{l}{\textit{Gamification}}\\
M1 & G & D1 Telemetry & T1-T4 & $\geq\!50\%$ LSG-CV\\
M2 & G & D1 Telemetry & T1-T4 & $\geq\!5$ days\\
M3 & G & LSG Ledger   & T1-T4 & $\geq\!60\%$ pts. earned\\
M4 & G & D1 Telemetry & T5     & $\geq\!2$ achiev./part.\\
M5 & G & GMS\textsuperscript{a} & T0, T3, T5 &
         $\Delta > 0$ LSG-CV vs.\ LSG-SV\\
\hline
\multicolumn{6}{l}{\textit{User Interface}}\\
M6 & I & SUS\textsuperscript{b} & T0, T3, T5 &
         $\geq\!70$ (good)\\
M7 & I & UEQ-S\textsuperscript{c} (8 items) & T0, T3, T5 &
         $\geq\!1.5$ per scale\\
M8 & I & Usability test & T1 & $\geq\!80\%$\\
M9 & I & D1 Telemetry   & T1 & $\leq\!5$ min.\\
M10& I & D1 Telemetry   & T1-T4 & $\leq\!5\%$\\
\hline
\multicolumn{6}{l}{\textit{Portal Commitment Compliance}}\\
M11 & C & Pipeline P2   & T1-T4 & $\geq\!70\%$ valid days\\
M12 & C & Mission engine & T1-T4 & $\geq\!80\%$ when index low\\
M13 & C & D1 Telemetry  & T1-T4 & $\geq\!40\%$ (opt-in)\\  
M14 & C & Observability & Continuous & p95 $\leq 250$ ms\\
M15 & C & Ledger audit  & Continuous & $100\%$ traceable\\
\hline
\multicolumn{6}{p{0.92\linewidth}}{\textbf{Metric} - M1: 7-day retention rate, M2: Mean valid-day streak, M3: Point redemption rate, M4: Achievements unlocked, M5: Self-determined motivation, M6: Perceived usability, M7: Experience quality, M8: Task completion rate, M9: Time to first action, M10: Exchange error rate, M11: Signal admissibility rate, M12: Mission activation rate, M13: Mission completion rate, M14: API gateway latency, M15: Rule-effect coherence.
\textbf{Instruments} -
\textsuperscript{a}Gamification Motivation
Scale~\cite{bocklage2019},
\textsuperscript{b}System Usability Scale~\cite{brooke1996},
\textsuperscript{c}User Experience Questionnaire
Short~\cite{laugwitz2008}.}
\end{tabular}
\end{table}

Metrics G (M1-M5) verify structural adherence and motivation \cite{seaborn2015}; metrics I (M6-M10) evaluate usability (SUS $\geq\!70$ \cite{bangor2009}, UEQ-S $\geq\!1.5$ \cite{laugwitz2008}); metrics C (M11-M15) guarantee pipeline integrity and technical latency. The LSG-CV vs.\ LSG-SV comparison on M5, M6, and M7 constitutes the empirical evidence for answering RQ3 (Sect.~\ref{sec:protocol}). These portal metrics are collected alongside the player-outcome battery detailed in Sect.~\ref{sec:protocol} (5.4), sharing the same T0-T5 schedule.

\subsection{Synchronization Architecture and Pipeline}

The LSG architecture implements real $\leftrightarrow$ virtual loops through five stages:

\begin{enumerate}
    \item \textbf{P1. Capture:} BLE sensors (smartbands, pedometers, sleep monitors) and digital cognitive tests via REST APIs.
    \item \textbf{P2. Preprocessing:} 7-day rolling window, outlier filtering, and conservative imputation (admissibility: $\geq\!70\%$ valid daily samples).
    \item \textbf{P3. Normalization:} Signals are normalized to $[0,1]$ using strategies F1-F4 (Sect.~3.4.1, L2), selected according to the clinical distribution of each signal.
    \item \textbf{P4. Aggregation:} Geometric mean of $\mathrm{IC}_{phys}$, $\mathrm{IC}_{ment}$ $\rightarrow$ $\mathrm{IC}_{LSG} \in [0,1]$, penalizing dimensional imbalance.
    \item \textbf{P5. Mapping:} The $\mathrm{IC}_{LSG}$ vector feeds the LSG-Core-API, which applies the R1-R6 catalogue; game events generate callbacks (virtual $\rightarrow$ real), closing the loop.
\end{enumerate}

\section{Game Mechanics Mapping Catalogue}
\label{sec:catalogue}

The catalogue translates $\mathrm{IC_{LSG}}$ index values into concrete game mechanics modifications (Table \ref{tab:rules}). Rules R1-R3 are efficiency rewards activated when indices sustain positive thresholds: R1 rewards sustained physical performance, R2 mental performance, and R3 activates a composite synergy bonus. Rules R4-R5 implement non-coercive safeguards (Principle P3): when indices fall critically low, optional recovery missions are generated in the LSG portal-never in-game penalties. Rule R6 activates a collective synergy scenario when both dimensions simultaneously exceed high thresholds, reinforcing cooperative play. All modifiers are temporary and bounded (Principle P2), and every rule-effect pair is fully traceable in the Ledger (Principle P1).

\begin{table}[ht]
\caption{LSG base mechanics mapping catalogue. R1-R3: Efficiency rewards; R4-R5: Non-coercive safeguards; R6: Synergy.}
\label{tab:rules}
\centering
\footnotesize
\renewcommand{\arraystretch}{1.1}
\begin{tabular}{p{0.8cm}p{2.4cm}p{3.2cm}p{2cm}p{1.7cm}p{1.4cm}}
\toprule
\textbf{ID} & \textbf{Condition} & \textbf{Game effect} &
\textbf{Type} & \textbf{Principle} & \textbf{Duration}\\
\midrule
R1 & $IC_{phys}\geq0.60$\newline (7\,days) &
     $+$10\,\% stamina;\newline $+$5\,\% movement\newline speed &
     Reward & P2 & 48\,h.\\
R2 & $IC_{ment}\geq0.55$ &
     $-$15\,\% cognitive\newline cooldown &
     Reward & P2 & 24\,h.\\
R3 & $IC_{LSG}\geq0.60$ &
     $+$5\,\% loot rate;\newline special mission &
     Synergy & P4 & 72\,h.\\
R4 & $IC_{phys}<0.40$\newline (3\,days) &
     Narrative recovery\newline mission in LSG\newline portal; soft\newline bonus cap &
     Safeguard & P3 & Until rec.\\
R5 & $IC_{ment}<0.35$ &
     Optional 5\,min.\newline cognitive mission\newline in LSG portal &
     Safeguard & P3 & Until compl.\\
R6 & $IC_{phys}\geq0.70$\newline $IC_{ment}\geq0.65$ &
     Synergy: special\newline narrative region\newline or cooperative\newline event &
     Collective\newline synergy & P4 & 96\,h.\\
\bottomrule
\end{tabular}
\end{table}

The specific threshold values in Table~\ref{tab:rules} are relative design decisions rather than absolute clinical cut-points: because $IC_{phys}$, $IC_{mental}$, and $IC_{LSG}$ are normalized to $[0,1]$ against each participant's own baseline (Sect.~3.4.1, L1-L2, established at $O_0$), 0.50 represents an individually-calibrated average state, and thresholds are set as offsets from that midpoint -- consistent with the \textit{challenge point} framework, where adaptive difficulty tracks an individual's own baseline rather than a population-wide standard \cite{guadagnoli2004}. Reward thresholds ($+0.05$ to $+0.20$) stay close to the midpoint to remain attainable through ordinary variation, while safeguard thresholds sit farther from it for the mental composite ($-0.15$) than the physical one ($-0.10$), since $IC_{mental}$ draws on shorter, less frequently sampled instruments than $IC_{phys}$'s continuous wearable telemetry -- a wider margin reduces the risk of a safeguard firing from measurement noise rather than a genuine decline, reinforced by the multi-day sustain windows (R1: 7 days; R4: 3 days) and the $\geq\!70\%$ admissibility floor (Sect.~3.4.1, L1). R6's higher bar ($\geq\!0.70$/$\geq\!0.65$) intentionally restricts collective synergy to a rare, aspirational state, consistent with Flow Theory's characterization of optimal experience as an infrequent peak \cite{csikszentmihalyi1990}. These values are design decisions pending empirical calibration during the pilot study (Sect.~\ref{sec:protocol}; Sect.~6.3). The catalogue is instantiated differently across the six commercial titles, reflecting the mechanical vocabulary of each genre (Table \ref{tab:games}). Signal-oriented games (\textit{Cities: Skylines}, \textit{Zero Sievert}) use a single dominant index to modulate core parameters; signal-plus-spatial combinations (\textit{Terraria}) require physical movement to generate signal. Temporal games (\textit{Stardew Valley}, \textit{Valheim}) accumulate signal between sessions,instrumentalizing Montola's temporal dimension most directly. \textit{Minecraft} is the only title operating on the group composite index, activating the social dimension of pervasivity. This diversity confirms that the R1-R6 rule catalogue is genre-agnostic: the same rules generate meaningfully different experiences according to each game's mechanical vocabulary.

\begin{table}[ht]
\caption{LSG instantiation across six commercial titles. Note how each game combines different pervasive dimensions.}
\label{tab:games}
\centering
\footnotesize
\renewcommand{\arraystretch}{1.15}
\begin{tabular}{p{2.4cm}p{2cm}p{4.5cm}p{2.7cm}}
\toprule
\textbf{Game} & \textbf{Genre} & \textbf{LSG Mechanics\newline
Adaptation} & \textbf{Dimensions}\\
\midrule
\textit{Stardew Valley} & Farm\newline sim. &
Physical activity unlocks\newline seasonal events; $IC_{ment}$\newline
modulates NPC\newline friendship decay &
Temporal + Signal\\
\textit{Cities: Skylines} & City\newline builder &
$IC_{ment}$ generates planning\newline efficiency bonuses;\newline
$IC_{phys}$ modulates\newline citizen happiness &
Signal (oriented)\\
\textit{Terraria} & Action\newline sandbox &
$IC_{phys}$ controls stamina;\newline $IC_{ment}$ influences\newline
crafting success &
Signal + Spatial\\
\textit{Valheim} & Survival\newline RPG &
$IC_{LSG}$ modulates weather\newline and biome events; low\newline
index activates\newline rest bonuses &
Temporal + Signal\\
\textit{Minecraft} & Cooperative\newline exploration &
Group composite index\newline triggers cooperative\newline
narrative events &
Social + Signal\\
\textit{Zero Sievert} & Tactical\newline RPG &
$I_{dm}$ directly modulates\newline tactical difficulty\newline
and loot quality &
Signal (oriented)\\
\bottomrule
\end{tabular}
\end{table}

\section{Study Protocol}
\label{sec:protocol}

This section describes the protocol designed to empirically validate the LSG framework introduced in Sect.~\ref{sec:framework} and calibrate the R1-R6 catalogue presented in Sect.~\ref{sec:catalogue}. The study pursues three research questions (Sect.~5.1) through a counterbalanced crossover quasi-experimental design that compares two within-participant conditions across the same six commercial titles: LSG-CV, the LSG-linked condition with full signal integration, and LSG-SV, the standard condition without signal integration (Sect.~5.2). Adult video game players from the Metropolitan Region of Chile are recruited as participants (Sect.~5.3), and physical, cognitive, and behavioural instruments (Sect.~5.4) feed the same normalization and aggregation pipeline described in Sect.~3.5, under full institutional ethical approval (Sect.~5.5). As stated in Sect.~1 and detailed in the Limitations (Sect.~6.3), this protocol has not yet been executed; the contribution of this section is the design of the validation study, not its results.

\subsection{Research Questions and Hypotheses}

The three research questions are:

\begin{itemize}
  \item \textbf{RQ1}: Does the LSG integration layer improve game-time self-regulation compared to playing without integration?
  \item \textbf{RQ2}: Does playing with LSG produce better well-being outcomes (physical activity, cognitive performance, perceived well-being) than playing without integration?
  \item \textbf{RQ3}: Do the five design principles predict player experience quality as measured by validated playability instruments \cite{gonzalezsanchez2014}?
\end{itemize}

\subsection{Experimental Design}

A counterbalanced crossover quasi-experimental design will be used (LSG-CV and LSG-SV conditions in counterbalanced order, separated by a 2-week washout period to minimize carry-over effects between conditions). Each condition lasts $\approx$10 weeks (total: $\approx$22 weeks, 5-6 months per participant), with six observations per condition: one pre-test ($O_0$), four intervention ($O_1$-$O_4$), and one post-test ($O_5$), matching the T0-T5 time-points of the portal measurement plan (Table~\ref{tab:metricas_portal}).

\subsection{Participants}

Target sample: $n\approx70$-80 adults (18-35 years), residents of the Metropolitan Region of Chile, video game players ($\geq\!3$ h/week) with a smartphone and stable internet connection who consent to wearing a non-invasive wearable. Exclusion criteria include diagnosed conditions affecting physical activity or cognition.

\subsection{Measurement Instruments}

Instruments are organized into two complementary batteries, administered at the same T0-T5 time-points across both conditions. The \textit{player-outcome battery} covers physical, cognitive, and behavioural dimensions, feeding RQ1-RQ2: IPAQ-SF ($I_{pf}$), Pittsburgh PSQI ($I_{gph}$), wearable sensor ($I_{pf}$, $I_{gph}$), adapted N-Back ($I_{mlp}$), decision simulator ($I_{dm}$), Ryff SPWB-29 (well-being), and LSG D1 telemetry (self-regulation). The \textit{portal-evaluation battery} covers the gamification, interface, and commitment-compliance metrics detailed in Table~\ref{tab:metricas_portal} (Sect.~3.4.1), including GMS, SUS, and UEQ-S at T0, T3, T5, feeding RQ3. Both batteries share the same LSG D1 telemetry pipeline (Sect.~3.5) and the same LSG-CV vs.\ LSG-SV comparison structure.

\subsection{Ethical Considerations}

The protocol has been approved by the Institutional Ethics Committee. Provisions include full pseudonymization (code \texttt{LSG-PXXX}), separate storage of identifying data, and full disclosure of mapping rules in the informed consent form, in direct application of Principle P1.

\section{Discussion}
\label{sec:discussion}

\subsection{LSG Positioning: Layer, Not Dimension}

The distinction between \textit{dimension} and \textit{transversal layer} carries theoretical consequences: a fourth dimension would imply that player signals break a new magic circle boundary, but physiological signals instead make the three existing domains sensitive to their inhabitant. Positioning LSG as a PEIL layer makes the contribution more precise without redefining the pervasive ontology: the new heuristic attributes it requires \cite{medina2021} are not additional dimensions, but quality attributes of the integration layer -- signal transparency, mapping fairness, safeguard non-coercivity, and signal-mechanic-narrative coherence.

\subsection{The LSG Portal and the Limits of Gamification}

LSG differs from conventional gamification \cite{deterding2011} at two levels: at the game level, mechanics act as a behavioural environment, not an incentive; at the portal level, structural gamification \cite{deci1985,seaborn2015} (achievements anchored to verified signals) sustains inter-session adherence. Metrics M3 and M5 (Table \ref{tab:metricas_portal}) will empirically verify this, per Hamari et al. \cite{hamari2014}.

\subsection{Limitations}

The LSG framework is at the design and formalization stage; data collection has not yet begun. The R1-R6 catalogue thresholds, the portal parameters (streaks, point values, challenge frequency), and the signal-pervasivity relationship by game genre all require empirical validation during the pilot study. The modding approach is viable for four games with robust ecosystems (\textit{Stardew Valley}, \textit{Terraria}, \textit{Minecraft}, \textit{Cities: Skylines}), but requires custom integration for \textit{Valheim} and \textit{Zero Sievert}.

\section{Conclusions}
\label{sec:conclusion}

This paper introduced LifeSync-Games (LSG) as a \textit{player experience integration layer} (PEIL) that transversally instruments Montola's three pervasive dimensions \cite{montola2005} through verified physiological and cognitive signals. LSG captures real-world signals (physical activity, sleep quality, memory, and decision speed) via non-invasive sensors and cognitive tests, normalizes them into $\mathrm{IC_{LSG}} \in [0,1]$, and maps them bidirectionally to game mechanics across six commercial titles through the LSG portal. Five contributions follow from this reconceptualization. LSG is conceptualized as a transversal layer that makes the spatial, temporal, and social dimensions responsive to the player's verified real state, rather than as an independent additional dimension; it introduces a gamified integration artifact, the LSG portal, with three functions, inter-session gamification, real $\rightarrow$ virtual point-for-mechanics exchange, and game-driven real-world missions; it proposes five HCI design principles grounded in SDT and Flow Theory, applicable to both the game and the LSG portal; it instantiates a mechanics mapping catalogue across six commercial video game genres; and it specifies a study protocol for empirical validation, already granted ethical approval. Together, these five contributions constitute a design and protocol proposal rather than an empirically validated system: the R1-R6 thresholds, the portal parameters, and the game mechanics mapping are theoretically grounded design decisions that still require the pilot study described in Sect.~\ref{sec:protocol} for calibration and validation (Sect.~6.3 details this scope boundary). By treating the player's verified physiological and cognitive state as the primary context variable, LSG achieves a level of personalization that location, time, or social context alone cannot provide, deepening the ontology of pervasive games without redefining it.

\subsection{Future Work}

The immediate next step is to execute the study protocol (Sect.~\ref{sec:protocol}; $n \approx 70\text{-}80$ participants; crossover quasi-experimental design); analyzing these data will calibrate the R1-R6 thresholds and portal parameters while addressing research questions RQ1-RQ3. In parallel, the research agenda will unfold along three lines. First, Medina-Medina et al.'s heuristic evaluation framework \cite{medina2021} will be extended with attributes specific to signal-driven pervasive games: signal transparency, mapping fairness, safeguard non-coercivity, and signal-mechanic-narrative coherence. Second, longitudinal adherence studies will evaluate whether the LSG portal's structural gamification sustains long-term intrinsic motivation, per SDT criteria \cite{deci1985,ryan2000}. Third, the framework will be generalized to new genres and platforms (e.g., mobile, virtual reality, online cooperative), exploring group $\mathrm{IC_{LSG}}$ metrics to deepen the social dimension of pervasivity. In the longer term, LSG establishes a research agenda at the intersection of pervasive game design, health HCI, and adaptive systems, with applications spanning clinical, educational, and active-lifestyle contexts.

\bibliographystyle{splncs04}
\bibliography{referencias}

@inproceedings{montola2005,
  author    = {Montola, M.},
  title     = {Exploring the edge of the magic circle: Defining pervasive games},
  booktitle = {Proceedings of DAC},
  volume    = {1966},
  pages     = {103},
  year      = {2005}
}

@article{medina2021,
  author    = {Medina-Medina, N. and Gallardo, J. and Cerezo, E. and
               Gutiérrez, F.~L. and Arango-López, J.},
  title     = {Hacia una propuesta de evaluación heurística de
               experiencias de juego pervasivas},
  journal   = {Interacción Revista Digital de AIPO},
  volume    = {2},
  number    = {2},
  pages     = {42--53},
  year      = {2021}
}

@article{arango2021geopgd,
  author    = {Arango-López, J. and Gutiérrez Vela, F.~L. and
               Collazos, C.~A. and Gallardo, J. and Moreira, F.},
  title     = {{GeoPGD}: Methodology for the design and development
               of geolocated pervasive games},
  journal   = {Universal Access in the Information Society},
  volume    = {20},
  number    = {3},
  pages     = {465--477},
  year      = {2021},
  doi       = {10.1007/s10209-020-00736-3}
}

@inproceedings{arango2017review,
  author    = {Arango-López, J. and Collazos, C.~A. and
               Vela, F.~L.~G. and Castillo, L.~F.},
  title     = {A systematic review of geolocated pervasive games:
               A perspective from game development methodologies,
               software metrics and linked open data},
  booktitle = {International Conference of Design, User Experience,
               and Usability (DUXU 2017)},
  series    = {LNCS},
  volume    = {10289},
  pages     = {335--346},
  publisher = {Springer},
  year      = {2017},
  doi       = {10.1007/978-3-319-58637-3_26}
}

@inproceedings{deterding2011,
  author    = {Deterding, S. and Dixon, D. and Khaled, R. and
               Nacke, L.},
  title     = {From game design elements to gamefulness:
               Defining gamification},
  booktitle = {Proceedings of the 15th International Academic
               MindTrek Conference (MindTrek '11)},
  pages     = {9--15},
  publisher = {ACM},
  year      = {2011},
  doi       = {10.1145/2181037.2181040}
}

@inproceedings{hamari2014,
  author    = {Hamari, J. and Koivisto, J. and Sarsa, H.},
  title     = {Does gamification work? {A} literature review of
               empirical studies on gamification},
  booktitle = {Proceedings of the 47th Hawaii International
               Conference on System Sciences (HICSS)},
  pages     = {3025--3034},
  publisher = {IEEE},
  year      = {2014},
  doi       = {10.1109/HICSS.2014.377}
}

@article{seaborn2015,
  author    = {Seaborn, K. and Fels, D.I.},
  title     = {Gamification in theory and action: A survey},
  journal   = {International Journal of Human-Computer Studies},
  volume    = {74},
  pages     = {14--31},
  year      = {2015},
  doi       = {10.1016/j.ijhcs.2014.09.006}
}

@article{nacke2017,
  author  = {Nacke, Lennart E. and Deterding, Sebastian},
  title   = {The maturing of gamification research},
  journal = {Computers in Human Behavior},
  volume  = {71},
  pages   = {450--454},
  year    = {2017},
  doi     = {10.1016/j.chb.2016.11.062}
}

@article{guadagnoli2004,
  author  = {Guadagnoli, Mark A. and Lee, Timothy D.},
  title   = {Challenge point: a framework for conceptualizing the effects of various practice conditions in motor learning},
  journal = {Journal of Motor Behavior},
  volume  = {36},
  number  = {2},
  pages   = {212--224},
  year    = {2004},
  doi     = {10.3200/JMBR.36.2.212-224}
}

@article{johnson2016,
  author    = {Johnson, D. and others},
  title     = {Gamification for health and wellbeing:
               A systematic review of the literature},
  journal   = {Internet Interventions},
  volume    = {6},
  pages     = {89--106},
  year      = {2016},
  doi       = {10.1016/j.invent.2016.10.002}
}

@article{zhao2024,
  author    = {Zhao, H. and others},
  title     = {Digital games and health promotion:
               A meta-analytic review},
  journal   = {Games for Health Journal},
  volume    = {13},
  number    = {1},
  pages     = {1--18},
  year      = {2024},
  doi       = {10.1089/g4h.2023.0102}
}

@article{deci1985,
  author    = {Deci, E.~L. and Ryan, R.~M.},
  title     = {The general causality orientations scale:
               Self-determination in personality},
  journal   = {Journal of Research in Personality},
  volume    = {19},
  number    = {2},
  pages     = {109--134},
  year      = {1985},
  doi       = {10.1016/0092-6566(85)90023-6}
}

@article{ryan2000,
  author    = {Ryan, R.~M. and Deci, E.~L.},
  title     = {Self-determination theory and the facilitation of
               intrinsic motivation, social development,
               and well-being},
  journal   = {American Psychologist},
  volume    = {55},
  number    = {1},
  pages     = {68--78},
  year      = {2000},
  doi       = {10.1037/0003-066X.55.1.68}
}

@article{ryan2006,
  author    = {Ryan, R.~M. and Rigby, C.~S. and Przybylski, A.},
  title     = {The motivational pull of video games:
               A self-determination theory approach},
  journal   = {Motivation and Emotion},
  volume    = {30},
  number    = {4},
  pages     = {344--360},
  year      = {2006},
  doi       = {10.1007/s11031-006-9051-8}
}

@book{csikszentmihalyi1990,
  author    = {Csikszentmihalyi, M.},
  title     = {Flow: The Psychology of Optimal Experience},
  publisher = {Harper \& Row},
  address   = {New York},
  year      = {1990}
}

@article{bangor2009,
  author    = {Bangor, A. and Kortum, P. and Miller, J.},
  title     = {Determining what individual SUS scores mean: Adding an
               adjective rating scale},
  journal   = {Journal of Usability Studies},
  volume    = {4},
  number    = {3},
  pages     = {114--123},
  year      = {2009}
}

@article{laugwitz2008,
  author    = {Laugwitz, B. and Held, T. and Schrepp, M.},
  title     = {Construction and evaluation of a User Experience
               Questionnaire},
  booktitle = {USAB 2008. LNCS},
  volume    = {5298},
  pages     = {63--76},
  publisher = {Springer},
  year      = {2008},
  doi       = {10.1007/978-3-540-89350-9_6}
}

@inproceedings{bocklage2019,
  author    = {Böcklage, A. and Siegmund, B. and Masuch, M.},
  title     = {Measuring gamification motivation: A validated scale for gamified systems},
  booktitle = {Proceedings of CHI Extended Abstracts},
  publisher = {ACM},
  year      = {2019},
  doi       = {10.1145/3290607.3313069}
}

@inproceedings{brooke1996,
  author    = {Brooke, J.},
  title     = {{SUS}: A quick and dirty usability scale},
  booktitle = {Usability Evaluation in Industry},
  pages     = {189--194},
  publisher = {Taylor \& Francis},
  year      = {1996}
}

@article{gonzalezsanchez2014,
  author    = {González-Sánchez, J.~L. and Gutiérrez-Vela, F.~L.},
  title     = {Assessing the player interaction experiences based
               on playability},
  journal   = {Entertainment Computing},
  volume    = {5},
  number    = {4},
  pages     = {259--267},
  year      = {2014},
  doi       = {10.1016/j.entcom.2014.08.001}
}

@article{mortazavi2024,
  author  = {Mortazavi, Fatemeh and Moradi, Hadi and Vahabie, Abdol-Hossein},
  title   = {Dynamic difficulty adjustment approaches in video games: a systematic literature review},
  journal = {Multimedia Tools and Applications},
  volume  = {83},
  pages   = {83227--83274},
  year    = {2024},
  doi     = {10.1007/s11042-024-18768-x}
}

@article{bontchev2016,
  author  = {Bontchev, Boyan},
  title   = {Adaptation in affective video games: A literature review},
  journal = {Cybernetics and Information Technologies},
  volume  = {16},
  number  = {3},
  pages   = {3--34},
  year    = {2016},
  doi     = {10.1515/cait-2016-0032}
}

\end{document}